\documentclass[fleqn,a4paper,12pt]{article}
\usepackage{amssymb}
\usepackage{amsmath}
\usepackage[dvips]{graphicx}
\usepackage{placeins}
\usepackage{lscape}
\usepackage{a4}
\usepackage{epsfig}

\numberwithin{equation}{section}

\newcommand{\beq}{\begin{equation}} \newcommand{\eeq}{\end{equation}}
\newcommand{\bea}{\begin{array}} \newcommand{\eea}{\end{array}}
\newcommand{\ri}{{\mathrm i}}

\catcode`@=11 \long
\def\@caption#1[#2]#3{\par\addcontentsline{\csname
ext@#1\endcsname}{#1} {\protect\numberline{\csname
the#1\endcsname}{\ignorespaces #2}} \begingroup \small
\@parboxrestore \@makecaption{\csname fnum@#1\endcsname}
{\ignorespaces #3}\par \endgroup} \catcode`@=12

\newcommand{\so}{\sigma_1}

\newcommand{\st}{\sigma_3}

\newcommand{\p}{\partial}

\catcode`@=11 \long
\def\@caption#1[#2]#3{\par\addcontentsline{\csname
ext@#1\endcsname}{#1} {\protect\numberline{\csname
the#1\endcsname}{\ignorespaces #2}} \begingroup \small
\@parboxrestore \@makecaption{\csname fnum@#1\endcsname}
{\ignorespaces #3}\par \endgroup} \catcode`@=12

\begin{document}
\allowdisplaybreaks
 \begin{titlepage}

 \vskip 2cm

\begin{center} {\Large\bf Matrix superpotentials and superintegrable
systems for arbitrary spin} \footnote{E-mail:
{\tt nikitin@imath.kiev.ua} } \vskip 3cm {\bf {A. G. Nikitin }
\vskip 5pt {\sl Institute of Mathematics, National Academy of
Sciences of Ukraine,\\ 3 Tereshchenkivs'ka Street, Kyiv-4, Ukraine,
01601\\}}
\end{center}
\vskip .5cm \rm
\begin{abstract}
A countable set of quantum superintegrable systems for arbitrary
spin is solved explicitly using tools of supersymmetric quantum
mechanics. It is shown that these
 systems (introduced by Pronko \cite{Pron2}) are special cases of models with shape
 invariant effective potentials that have recently been classified
 in \cite{Nik1} and \cite{Nik2}.

\end{abstract}
\end{titlepage}

\section{Introduction\label{intro}}

  Exactly solvable problems of quantum mechanics are very interesting
  and important. Their solutions can be found in a straightforward way
  free of uncertainties caused  by  perturbation methods.  The very
  existence of these solutions  is usually connected with  symmetries
  which are very attractive subjects by themselves. Moreover, exact
  solutions present convenient bases for expansion of solutions of other
  problems.

  Another nice property of some quantum mechanical systems is called
  superintegrability. The system is called maximally superintegrable if there exist a
  sufficient number of algebraically independent operators commuting with
  the Hamiltonian. This number should be equal to $2n-2$ for the system with
  $n$ degrees of freedom, and $n-1$ of these operators should commute
  amongst each other.

  Many of exactly solvable systems are maximally superintegrable, see,
  e.g.,  \cite{wint}, and wise versa. A perfect example of quantum mechanical system which is both maximally
  superintegrable and exactly solvable is the non-relativistic Hydrogen
  atom (NRHA). The concept of superintegrability had been extended to the case of quantum systems which include spin \cite{wint2}.  However, the completed relations between the exact solvability and  superintegrability are not clear yet, especially for the higher order integrals of motion and for the systems with $n>2$.

An alternative point of view on the exactly solvable QM systems was
created by Gendenshtein \cite{Gen} who had shown that in many cases
such systems admit supersymmetry with shape invariance and can be
easily solved using tools of supersymmetric QM. In particular it is
true for the NRHA.

A consistent maximally superintegrable and exactly solvable problem  was discovered
by Pron'ko and Stroganov \cite{Pron}. The related quantum mechanical system
includes a magnetic dipole with spin $\frac12$ (neutron) moving in
the field of a stright line current. Like the NRHA, this problem
admits the  hidden symmetry which is more extended then its
geometric symmetry \cite{Pron}. It is supersymmetric also and can be
simply solved using the shape invariance of its effective potential
\cite{Vor}, \cite{Gol}.

The physical and mathematical aspects of the Pron'ko-Stroganov
\cite{Pron} (PS) model was multiple investigated. Its experimental
realizability was discussed in papers \cite{Blum1} and \cite{Blum2},
its symmetries and supersymmetries was studied in \cite{mota} and
\cite{rodric},  enhanced analysis of supersymmetric aspects of this
model can be found in \cite{Ioffe}, etc. Recently the solvable
relativistic version of the PS model was found \cite{nika} which,
however, includes a  more complicated external field than in the
non-relativistic case.

Whenever there exist a good model for a particle of spin $\frac12$,
there is a natural desire to generalize it for  higher spins. In the
case of the PS model the time needed to realize this idea was thirty
long years.  The main hidden dangers in this generalization were
connected with the fact that the direct change of the Pauli
interaction term present in the PS model by the analogous term for
higher spin leads to the system which is not exactly solvable
  and its finite trajectories are not closed \cite{Gol2}.

In paper \cite{Pron2} exactly solvable generalizations of the PS model
 to the case of arbitrary spin have been
formulated. The price paid for this progress was the essential
complication of the Pauli interaction term present in the PS model.
However, there are good physical arguments for such a complication in the case of spin $s>\frac12$,
in both the non-relativistic \cite{Pron2} and relativistic
\cite{Beckers} approaches.

A natural question arises whether the models proposed
in \cite{Pron2} can be effectively integrated using the tools of
SUSY quantum mechanics, like it has been done \cite{Gol} for the PS
model for neutral particle of spin $\frac12$. This question is
especially provocative for us since in the recent papers \cite{Nik1}
and \cite{Nik2} an effective classification of shape invariant
matrix potentials was presented. Thus if the matrix potentials found
in \cite{Pron2} are shape invariant they should be nothing but
particular cases obtained by this classification.

In the present paper the supersymmetric aspects of the maximally
superintegrable systems for arbitrary spin (proposed in
\cite{Pron2}) are discussed. It is shown that these systems are
shape invariant and can be easily integrated using the tools of SUSY
quantum mechanics. Using the results of paper \cite{Nik1} we
construct exact solutions of the mentioned systems directly for
arbitrary spin. In addition, a more straightforward and refined
formulation of these systems is proposed together with the proper
physical interpretation.

\section{Hidden symmetry of the PS model}

The PS model is based on the following version of the
Schrodinger-Pauli Hamiltonian:
\begin{gather}{\cal H}=\frac{p_x^2+p_y^2}{2m}+\lambda\frac{S_xy-S_yx}{r^2}
\label{ps2}\end{gather} which is the Hamiltonian of a neutral spinor
anomalously interacting with the magnetic field generated by a
straight  line current directed along the $z$ coordinate axis.  Here
$p_x=-\ri \frac{\partial}{\partial x}, p_y=-\ri
\frac{\partial}{\partial y},\ r^2=x^2+y^2$; $S_x=\frac12\sigma_1$ and
$S_y=\frac12\sigma_2$ are matrices of spin $\frac12$, $\sigma_1$ and
$\sigma_2$ are Pauli matrices, $\lambda$ is the integrated coupling
constant.

The last term in (\ref{ps2}) is the Pauli interaction term
$\lambda{\bf S}\cdot{\bf H}$ where the magnetic field ${\bf H}$ has
the following components which we write ignoring the constant multiplier included into the parameter $\lambda$:
\begin{gather}H_x\sim\frac{y}{r^2},\quad H_y~\sim-\frac{x}{r^2}, \quad
H_z=0.\label{ps2a}\end{gather}

 Hamiltonian (\ref{ps2}) is invariant
w.r.t. rotations around the $z$-axis since it commutes with the $z$
component of the total angular momentum
\begin{gather}\label{J3}J_z=x_1p_2-x_2p_1+S_z\end{gather}
where $S_z=\frac12\sigma_3$. In addition, it admits two more
constants of motion \cite{Pron}, namely:
\begin{gather}A_x=\frac12(J_zp_x+p_xJ_z)+\frac{m}{r}\mu({\bf n})y,
\quad A_y=\frac12(J_zp_y+p_yJ_z)-\frac{m}{r}\mu({\bf
n})x\label{AA}\end{gather} where $\mu({\bf
n})=\lambda({S_xn_y-S_yn_x})$, $n_x=\frac{x}r, n_y=\frac{y}r, $ and
${\bf n}=(n_x,n_y)$.  Operators (\ref{J3}) and (\ref{AA}) commute
with $\cal H$ and satisfy the following commutation relations:
\begin{gather}[J_z,A_x]=\ri A_y,\quad [J_z,A_y]=-\ri A_x,\quad
[A_x,A_y]=-2\ri m J_z\cal H.\label{AAJ}\end{gather} In other words,
operators (\ref{AA}) and (\ref{AAJ}) represent algebra o(3) on the
spaces of eigenvectors of $\cal H$ with negative eigenvalues  and  cause the degeneration of
these eigenvalues. And this hidden symmetry makes the PS model maximally superintegrable and exactly solvable.

It is interesting to note that up to unitary equivalence operator
(\ref{ps4}) is the only plane Hamiltonian for neutral particle of
spin $\frac12$ interacting with an external field, which commutes
with $J_3$ and admits a hidden symmetry w.r.t. algebra o(3) for negative energies and algebra o(1,2) for positive energies. This
statement in fact was proven in paper \cite{Pron} where, however, a
more general Hamiltonian has been proposed, namely:
\begin{gather}\hat {\cal H}=\frac{p_x^2+p_y^2}{2m}+a\frac{S_xy-S_yx}{r^2}+
b\frac{S_xx+S_yy}{r^2}\label{ps3}\end{gather}

Operator (\ref{ps3}) does commute with (\ref{AA}) and (\ref{AAJ}) and includes two arbitrary
parameters $a$ and $b$.
But it is unitary equivalent to (\ref{ps4}) since
\begin{gather}\hat {\cal H}=U{\cal H}U^\dag\label{T1}\quad \text{with}\quad
U=\cos\theta+\ri S_z\sin\theta\end{gather} where
$S_z=\frac12\sigma_3$  and $\theta$ is a real parameter such that
$a=\lambda\cos\theta$ and $b=\lambda\sin\theta$.

\section{SUSY aspects of PS model}\label{psm}

Let us apply transformation (\ref{T1}). For our purposes it is
convenient to set $\theta=\pi/2$ and obtain the following
Hamiltonian which is unitary equivalent to (\ref{ps2}):
\begin{equation}\hat {\cal
H}=\frac{p_x^2+p_y^2}{2m}+
\lambda\frac{S_xx+S_yy}{r^2}.\label{ps4}\end{equation} Then we
consider the eigenvalue problem for Hamiltonian (\ref{ps4}):
\begin{gather}\hat{\cal H}\psi={\cal E}\psi\label{ps1}\end{gather}
where $\psi=\psi(x,y) $ is a two-component wave function.
This function is supposed to be square integrable and vanish at $x=y=0$.

  Introducing the polar coordinates
  \begin{gather}\label{polco}x=r\text{cos}\theta, \quad
  y=r\text{sin}\theta\end{gather}  and expanding $\psi$ via eigenfunctions of the
angular momentum operator $J_{z}$: \beq\label{s1}\psi=  C_\kappa
\psi_\kappa,\ \ \ \psi_\kappa=\frac{1}{\sqrt{r}}
\left(\bea{c}\texttt{exp}(\texttt{i}(\kappa-\frac12)\theta)\phi_1\\
\epsilon\texttt{exp}(\texttt{i}(\kappa+\frac12)\theta)\phi_2\eea\right)\eeq
where $C_\kappa$ are constants, $\phi_1$ and $\phi_2$ are functions
of $r$ and summation is imposed over the repeated indices
$\kappa=\pm\frac12,\pm\frac32,\pm\frac52\cdots$, equation
(\ref{ps1}) is reduced to the following system of decoupled
equations for $\psi_\kappa$: \beq\label{ps5}{\cal
H}_\kappa\psi_\kappa\equiv\left(-\frac{\partial^2}{\partial
r^2}+\kappa(\kappa-\sigma_3) \frac{1}{r^2}+\sigma_1\frac{\tilde
\lambda} {{r}}\right)\psi_\kappa={\cal {\tilde E}}\psi_\kappa\eeq
where ${\cal {\tilde E}}={2m\cal E},\ \tilde \lambda=2m\lambda$.

Equation (\ref{ps4}) and its reduced version (\ref{ps5}) are
invariant with respect to the space reflection $x\to x, y\to -y$
provided $\psi(x,y)$ and $\psi_k(r)$ cotransform as
\begin{gather}\psi(x,y)\to
\Lambda\psi(x,-y),\quad
\psi_k(r)\to\Lambda\psi_{-k}(r)\label{ps71}\end{gather}
where $\Lambda=\sigma_1$.

Transformations (\ref{ps71}) commute with Hamiltonians $\cal H$ and
${\cal H}_\kappa$, anticommute with $J_3$ and are invertible. Thus
we can restrict ourselves to equation (\ref{ps5}) for positive half
integer $k$. Then solutions for negative $k$ can be obtained using
(\ref{ps71}).

 Hamiltonian ${\cal H}_\kappa$ can be factorized as
\beq\label{ps6}{\cal H}_\kappa=a_\kappa^+a_\kappa+c_\kappa\eeq where
\begin{gather}\label{a+-}a_\kappa=\frac{\partial}{\partial r}+W_\kappa,\  \ a_\kappa^+=-
\frac{\partial}{\partial r}+W_\kappa,\ \
c_\kappa=-\frac{\tilde\lambda^2}{(2k+1)^2}\end{gather} and $W$ is the {\it
matrix superpotential}
\beq\label{s4}W_\kappa=\frac{1}{2r}\sigma_3-\frac{\tilde\lambda}{2\kappa+1}\sigma_1-
\frac{\left(\kappa+\frac12\right)}{r}.\eeq

One more important  property of  ${\cal H}_\kappa$ is its shape
invariance, i.e.,  its superpartner ${\cal H}^+_\kappa$ is equal to
${\cal H}_{\kappa+1}$ up to a constant term:
\[{\cal H}_\kappa^+=a_\kappa^-a_\kappa^++c_\kappa=
-\frac{\partial^2}{\partial r^2}+(\kappa+1)(\kappa+1-\st)
\frac{1}{r^2}+\so\frac{\tilde\lambda} r={\cal H}_{\kappa+1}+C_\kappa\] where
$C_\kappa=c_\kappa-c_{\kappa+1}$. Thus equation (\ref{ps5}) can be
solved using the standard tools of SSQM. Namely, the ground state
vector is defined as a square integrable solution of the first order
equation
\begin{gather}a_\kappa^-\psi_{\kappa,0}(r)\equiv\left( \frac{\p}{\p
  r}+W_\kappa\right)\psi_{\kappa,0}(r)=0\label{ps7}\end{gather} and
  thus it has the following components
\begin{gather}\label{ps8}\phi_1=u^{\kappa+1}
K_{1}(u), \quad \phi_2=u^{\kappa+1} K_{0}(u)
\end{gather}
where $K_0$ and $K_1$ are the modified Bessel functions,
$u=\frac{\tilde\lambda r}{2\kappa+1}$.

 In view of
(\ref{ps6}) function $\psi_{\kappa,0}(r)$ solves also equation
(\ref{ps5}) with ${\cal {\tilde
E}}=c_\kappa=-\frac{\tilde\lambda^2}{(2k+1)^2}$.
Solutions
  which correspond to $n^{th}$ exited state
  can be represented as
\begin{gather}\label{ps9}\psi_{\kappa,n}(r)=
a_\kappa^+a_{\kappa+1}^+ \cdots
a_{\kappa+n-1}^+\psi_{\kappa+n,0}(x).
\end{gather} The corresponding eigenvalue $\tilde{\cal E}_n$ is given by the following formula:
\begin{equation}
\label{eigenvalues} \tilde{\cal E}_n
=\sum\limits_{i=0}^{n-1}C_{\kappa+i}=-\frac{\tilde\lambda^2}{(2\kappa+2n+1)^2}.
\end{equation}


\section{Integrable models for neutral vector bosons}\label{VB}
The PS model considered in the previous section describes the
interaction of a neutral particle of spin $\frac12$ with the field
of straight line current. How we can generalize Hamiltonian
(\ref{ps4}) (or the initial Hamiltonian (\ref{ps2}) which is unitary
equivalent to (\ref{ps4})) to the case of higher spins? The standard
idea is simple to change in (\ref{ps4}) matrices
$S_\alpha=\frac12\sigma_\alpha$ by matrices
 of higher spin. However, in this way we obtain a model
 which is neither exactly solvable \cite{Gol2} nor  superintegrable.

 To obtain a  superintegrable analogues of (\ref{ps2}) for
 higher spins it is necessary to make a rather non-trivial
 generalization of the Pauli interaction term in the Hamiltonian
 (\ref{ps4}), including multipole magnetic interactions
 \cite{Pron2}.

 In this section we present an analogue of the PS model for particle
 of spin $1$.
 This model is both superintegrable and shape invariant.
 It is based on the following Hamiltonian
\begin{gather}{\cal H}_s=\frac{p_x^2+p_y^2}{2m}+\frac1{r}\mu_s(
{\bf n}) \label{ps21}\end{gather} where
\begin{gather}\label{mus} \mu_s( {\bf n})=\mu_1( {\bf n})=\mu (2({\bf
S}\times{\bf n})^2-1) +\lambda (2({\bf S}\cdot{\bf
n})^2-1).\end{gather} Here $\mu$ and $\lambda$ are arbitrary real
parameters,  ${\bf S}\cdot{\bf n}=S_xn_x+S_yn_y$ and ${\bf
S}\times{\bf n}=S_xn_y-S_yn_x$, $S_x$ and $S_y$ are matrices of spin
1:
\begin{gather} S_x=\frac1{\sqrt2}\begin{pmatrix}0&1&0\\1&0&1\\0&1&0\end{pmatrix},
\quad S_y=
\frac\ri{\sqrt2}\begin{pmatrix}0&-1&0\\1&0&-1\\0&1&0\end{pmatrix},
\quad
S_z=\begin{pmatrix}1&0&0\\0&0&0\\0&0&-1\end{pmatrix}.\label{SPIN1}
\end{gather}

It is the Hamiltonian defined by equations (4.1) and (4.2) that
generalized  (2.1) for the case of spin one. Matrix (\ref{mus})
depends on ${\bf n}=(\frac{\bf x}{r},\frac{\bf y}{r})$ and satisfies
the following conditions:
\begin{gather}[\mu_s({\bf n}),J_z]=0, \label{condi1}\\
 \mu_s({\bf n})S_z+S_z\mu_s({\bf
n})=0.\label{condi2}\end{gather} Thus Hamiltonian  (\ref{ps21})
commutes with operators (\ref{J3})
 and (\ref{AA}) where  $\mu({\bf n})\to\mu_1({\bf n})$ and $S_z$ is the matrix given in (\ref{SPIN1}).
 So this Hamiltonian admits dynamical symmetry w.r.t. algebra
 o(3) realized on states with negative energies.

 It is possible to show that, up to unitary equivalence, equations (\ref{ps21}), (\ref{mus}) represent the most general form of plane Hamiltonian for neutral particle of spin 1, which admits this symmetry, see section 7.
We will see that, in addition, this Hamiltonian is  shape invariant.
 Let us discuss its physical
content.

The physical sense of the PS model for spin $\frac12$ is absolutely
clear. The related Hamiltonian (\ref{ps21}) includes the Pauli term
$\sim \bf S\cdot H$ and corresponds to the interaction of a neutral
particle (having a non-trivial dipole moment) with the field of the
constant and straight line current.

Physical content of superintegrable models for higher spins is much
more sophisticated.  The interaction term of Hamiltonian
(\ref{ps21}) has nothing to do with the Pauli term and needs another
interpretation.

Let us consider in more detail the case $\mu=0, \ \lambda=\omega>0$. The
corresponding
 Hamiltonian (\ref{ps21}) takes the following form:
 \begin{gather}{\cal H}_1=\frac{p_x^2+p_y^2}{2m}+
 \omega\left(\frac{2({\bf S}\cdot{\bf x})^2}{r^3}-\frac1{r}\right).
 \label{IHH}\end{gather}
and can be represented as
\begin{gather}\label{rep2}{\cal H}_1=\frac{p_{x_1}^2+p_{x_2}^2}{2m}+
\omega Q_{ab}\frac{\p E_a}{\p x_b}+\frac\omega3
\texttt{div} {\bf E}\end{gather} where
\begin{gather*}Q_{ab}=S_aS_b+S_bS_a-\frac{2}3s(s+1)\delta_{ab}\end{gather*}
is the tensor of quadruple interaction,
  \begin{gather}\label{E}{\bf E}=\left(\frac{ x_1}{r}, \frac{ x_2}{r},
  0\right),
  \end{gather} and the temporary
  notations
   $x=x_1, y=x_2$ are used.

  In accordance with (\ref{rep2}) ${\cal H}_1$ can be interpreted as a Hamiltonian
  of spin-one particle which has neither minimal nor dipole interaction
with the external field. However, this particle is supposed to have
a quadruple  and Darwin interaction, the latest is represented by
the last term in (\ref{rep2}).

A more delicate question is related to the physical realizability of
the vector field (\ref{E}) included into the interaction terms.
If  we suppose
that it is the classical Maxwell electric field, the corresponding charge density should be proportional to  $\texttt{div}{\bf
E}=\frac1r$. Such charge density is seemed to be hardly realized
experimentally. However,
vector (\ref{E})
perfectly solves field equations of generalized Maxwell
electrodynamics modified by presence of Chern-Simons term, and
also equations of axion electrodynamics  with trivial current
and charge densities \cite{NKU}.

Let us  present one more expression of the interaction term via
physical fields. Setting $\lambda=0,\ \mu=\omega>0$  we can rewrite
Hamiltonian (\ref{ps21}) in the following form:
\begin{gather}\label{rep1}\hat{\cal H}_1=\frac{p_{x}^2+p_y^2}{2m}+ \omega\frac{2({\bf S}\cdot{\bf
H})^2-{\bf H}^2}{|\bf H|},\end{gather} where ${\bf H}$ is the vector
of magnetic field whose components are defined in equation
(\ref{ps2a}).

Notice that Hamiltonians  (\ref{IHH}) and (\ref{rep1}) are unitary
equivalent. Namely,
\begin{gather}{\cal H}_s=U\hat{\cal H}_sU^\dag,\quad
U=\exp\left(\frac{\ri\pi}2S_z\right)\label{trans2}\end{gather} where
$s=1$.

The last term in (\ref{rep1}) represents a nonlinear interaction
with the well defined external magnetic field (\ref{ps2a}). This
field solves the Maxwell equations with a constant straight line
current.

Let us remind that a nonlinear generalization of the Pauli
interaction is requested also in the relativistic description of
spin-one particle interacting with the constant magnetic field
\cite{Beckers}.

\section{Superintegrable models for spin $\frac32$}\label{s3/2}

 Let us discuss an exactly solvable model for particle of spin $\frac32$.
  The corresponding Hamiltonian has the generic form (\ref{ps21})
 with
\begin{gather}\label{mu32} \mu_s({\bf
n})=\left(\nu+\mu S_3^2\right)\left(7{\bf S\times{\bf n}}-4({\bf
S\times{\bf n}})^3\right).\end{gather}
 Here $\mu$ and $\nu$ are arbitrary parameters, $S_x, S_y$ and $S_z$ are the
 $4\times4$  matrices of spin $\frac32$ which can be chosen in the following
 form:
\begin{gather}\begin{split}&S_x=\frac12\begin{pmatrix}0&\sqrt{3}&0&0\\\sqrt{3}&0&2&0\\0&2&0&\sqrt{3}\\
  0&0&\sqrt{3}&0
  \end{pmatrix},\quad S_y=\frac{\ri}2\begin{pmatrix}0&-\sqrt{3}&0&0\\\sqrt{3}&0&-2&0\\0&2&0&-\sqrt{3}\\
  0&0&\sqrt{3}&0
  \end{pmatrix},\\& \qquad \qquad \qquad \qquad  S_z=\frac12\begin{pmatrix}3&0&0&0\\0&1&0&0\\0&0&-1&0\\
  0&0&0&-3
  \end{pmatrix}.\end{split}\label{3/2}\end{gather}

Matrix (\ref{mu32}) satisfies conditions (\ref{condi1}) and (\ref{condi2}) which are
necessary and sufficient for existence of constants of motion
(\ref{J3}) and (\ref{AA}) \cite{Pron2}. Thus the system whose Hamiltonian is given
by equations (\ref{ps21}) and (\ref{mu32}) actually is maximally superintegrable.

The interaction term of the considered Hamiltonian can be represented
in terms of external fields.
In the case $\mu=0, \ \nu=\frac\lambda3\neq0$ we have:
\begin{gather}{\cal
H}=\frac{p_x^2+p_y^2}{2m}+
  \omega\left({\bf S}\cdot{\bf H}-
  \frac47\frac{({\bf S}\cdot{\bf H})^3}{{\bf H}^2}
  \right)\label{IHIH}\end{gather}
 where $\omega=\frac{7\lambda}3$, ${\bf H}$ is the vector of magnetic
field whose components are defined in equation (\ref{ps2a}) and $\bf
S$ is the spin $\frac32$ vector with components (\ref{3/2}).

  In addition to the standard Pauli term $\omega{\bf S}\cdot{\bf H}$
  Hamiltonian (\ref{IHIH})
  includes the additional  interaction
  $\sim \frac{({\bf S}\cdot{\bf H})^3}{{\bf H}^2}$ which is
  non-linear in magnetic field.

  Alternatively, for $\mu=-\frac43\nu$ we obtain the following
  representation:
\begin{gather}{\cal
H}=\frac{p_{x_1}^2+p_{x_2}^2}{2m}-
  \nu Q_{abc}\frac{\p^2 H_a}{\p x_b\p x_c}\label{IHIHI}\end{gather}
where
\[Q_{abc}=\sum_{P(a,b,c)}\left(S_aS_bS_c-
\frac{7}4S_a\delta_{bc}\right)\] is the octuple interaction tensor
(the summation is imposed over all possible permutations of indices
$a, b$ and $c$), and ${\bf H}=\left(x_2\ln r,-x_1 \ln r, 0\right).$

The vector field $\bf H$ perfectly solves equations of
the axion electrodynamics \cite{NKU}. On the other hand, being treated as the
classical magnetic field it requires a current whose density grows
algorithmically with growing of $r$.

\section{Superintegrable systems for arbitrary spin}\label{AS}

 In paper \cite{Pron2} the generalization of Hamiltonian (\ref{ps2})
 for arbitrary spin $s$ was proposed, which  admit dynamical symmetry w.r.t.
 algebra  o(3). This Hamiltonian has the generic form (\ref{ps21}) where $\mu_s({\bf n})$ is a
$(2s+1)\times(2s+1)$ dimensional matrix depending on ${\bf
n}=(\frac{\bf x}{r},\frac{\bf y}{r})$ and satisfying conditions
(\ref{condi1}), (\ref{condi2}) with $S_z$ being the $z$
  component of spin
$s$ vector, i.e., the  matrix
\begin{gather}\label{S3}S_z=\text{diag}(s,s-1,s-2,...-s).\end{gather}

It can be verified by direct calculations that if conditions
(\ref{condi1}) and (\ref{condi2}) are satisfied then Hamiltonian (\ref{ps21}) commutes
with operators (\ref{J3}) and (\ref{AA}) (where $S_z$ is matrix
(\ref{S3}) and $\mu(\bf n)\to \mu_s(\bf n)$), and these operators do
satisfy relations (\ref{AAJ}).

We shall refine and
extend  the results of paper \cite{Pron2}. First, we present a
straightforward  formulation of Hamiltonians (\ref{ps21}) for
arbitrary spin. Secondly, we prove that the number of arbitrary
parameters present in the models found in \cite{Pron2} can be
effectively reduced applying a unitary transformation.
At the third place  the shape invariance of
 Hamiltonians (\ref{ps21}) will be proven and the corresponding
eigenvalue problems will be solved algebraically.

Let us find Hamiltonians (\ref{ps21}) for
arbitrary spin. The condition (\ref{condi1}) is satisfied
iff $\mu_s({\bf n})$ is a function of ${\bf S}\cdot{\bf
n}=S_xn_x+S_yn_y, \  {\bf S}\times{\bf n}=S_xn_y-S_yn_x$ and $S_z$.
At the first step we restrict ourselves to
 matrices $\mu_s({\bf n})$ which are polynomials in  ${\bf S}\cdot{\bf
n}$.

It is convenient to represent
matrix $\mu_s({\bf n})$ in the following form:
\begin{gather}\label{mu} \mu_s({\bf n})=\sum_{\nu=-s}^s
c_\nu\Lambda_\nu\end{gather} where $c_\nu$ are unknown coefficients
and $\Lambda_\nu$ are projectors onto the eigenspaces of matrix
${\bf S}\cdot{\bf
n}$ corresponding to the eigenvalue $\nu$ ($\nu,
\nu'=s,s-1,...-s$):
\begin{gather}\label{Lnu}\Lambda_\nu=\prod_{\nu'\neq\nu}
\frac{{\bf S}\cdot{\bf
n}-\nu'}{\nu-\nu'}.\end{gather}

Matrix (\ref{mu}) by construction satisfies condition
(\ref{condi1}). Substituting (\ref{mu}) into (\ref{condi2}) and
using the identities \[S_z\mu_s({\bf n})+ \mu_s({\bf
n})S_z=2S_z\mu_s({\bf n})+[\mu_s({\bf n}),S_z]\] and \cite{FGN}\footnote{Formula (\ref{Nik1}) is a particular case of equation (A.2) in \cite{FGN} corresponding to $\frac{ p_1}{p}={n_x}, \frac{ p_2}{p}={n_y}, \frac{ p_3}{p}=0, S_{23}=S_x, S_{31}=S_y$.
This formula is also a consequence of  equation (7.18) from book \cite{FN} }
\begin{gather}\label{Nik1}[\Lambda_\nu,S_z]=\frac12S_z(2\Lambda_\nu-
\Lambda_{\nu+1}-\Lambda_{\nu-1})+\frac{\ri}2\left(n_xS_y-n_yS_x\right)
\left(\Lambda_{\nu+1}-\Lambda_{\nu-1}\right)\end{gather} we easily
find the following condition for coefficients $c_\nu$ (note that projectors $\Lambda_\nu$ are orthogonal and  linearly independent):
\begin{gather}c_\nu=-c_{\nu-1},\quad
\nu=s,s-1,...,1-s.\label{nunu}\end{gather}

In accordance with (\ref{mu}) and (\ref{nunu}) the general
expression for $\mu_s({\bf n})$ can be given by the following
equation:
\begin{gather}\mu_s({\bf n})=\lambda\sum_\nu(-1)^{[\nu]}
\Lambda_\nu\label{i}\end{gather}
where $[\nu]$ is the entire part of $\nu$, $\nu=s, s-1,..., -s$.

Equations (\ref{ps21}), (\ref{i}) present Hamiltonians for arbitrary
spin $s$ admitting dynamical symmetry w.r.t. algebra o(3) for states with negative eigenvalues.
Potentials (\ref{i}) are defined up to arbitrary parameter $\lambda$
which can be associated with the coupling constant.

However in accordance with the results of paper \cite{Pron2} the
general solution of the determining equations (\ref{condi1}), (\ref{condi2}) depends
on $2s+1$ arbitrary real parameters. This statement is in accordance
with the fact that  there exist exactly $2s+1$ linearly independent
matrices anticommuting with $S_z$.

But our analysis admits a straightforward and simple extension to
the generic case. Indeed, equations (\ref{condi1}), (\ref{condi2}) are invariant w.r.t.
multiplying $\mu_s({\bf n})$ by arbitrary power of matrix $S_z$.
Thus, starting with (\ref{i})  we can immediately construct
$2s+1$-parametrical solutions of this equation:
\begin{gather}\label{GC}{\tilde\mu}_s({\bf n})=
\sum_{\nu\geq0}(b_\nu \tilde B_\nu+\ri d_\nu \tilde C_\nu)\mu_s({\bf
n})\end{gather} where $\mu_s({\bf n})$ are matrices (\ref{i}),
$b_\nu$ and $d_\nu$ are arbitrary (real) parameters, $\tilde B_\nu$
and $\tilde C_\nu$ are projectors, polynomial in $S_z$:
\begin{gather*}\tilde B_\nu=\prod_{\nu'\neq\nu}\frac{S_z^2-\nu'^2}{\nu^2-\nu'^2},
\quad \tilde
C_\nu=\frac{S_z}{\nu}\prod_{\nu'\neq\nu}\frac{S_z^2-\nu'^2}{\nu^2-\nu'^2},\quad
\nu, \nu'=s,s-1,s-2,...,\quad
 \nu, \nu'\geq0.\end{gather*}
Notice that $C_\nu$ includes only odd powers of $S_z$ and so they
anticommute with  $\mu_s({\bf n})$. Thus matrix ${\tilde\mu}_s({\bf
n})$ is hermitian.

 Using the orthogonality relations
\[\tilde B_\mu \tilde B_\nu=\tilde C_\mu \tilde C_\nu=
\delta_{\mu\nu} \tilde B_\nu,\quad \tilde C_\mu \tilde B_\nu=\delta_{\mu\nu}\tilde C_\nu\]
it is easy to prove that all terms in sums (\ref{GC}) are linearly
independent.  Thus matrix ${\hat\mu}_s({\bf n})$ includes exactly
$2s+1$ essential parameters and so it should be equivalent to the
analogous matrix found in \cite{Pron2}, see equation (29) there.
However the number of these parameters can be reduced using the
unitary transformation
\begin{gather}\label{T2}{\tilde\mu}_s({\bf n})\to
U {\tilde\mu}_s({\bf n}) U^\dag=
\sum_{\nu\geq0}{\lambda}_\nu B_\nu\mu_s({\bf
n})\equiv{\hat\mu}_s({\bf n})\end{gather} where
${\lambda}_\nu=\sqrt{b_\nu^2+d_\nu^2}$ and
$U=\sum_{\nu\geq0}(\cos\theta_\nu B_\nu+\ri\sin \theta_\nu C_\nu)$.
Taking into account that $\mu_s({\bf n})$ commutes with $B_\nu$ and
anticommutes with $C_\nu$ it is easy to make sure that relation
(\ref{T2}) is true provided parameters $\theta_\nu$ satisfy the
conditions $b_\nu=\lambda_\nu\cos\theta_\nu $ and
$d_\nu=\lambda_\nu\sin\theta_\nu $.

It follows from (\ref{T2}) that, up to unitary equivalence, matrix
${\hat\mu}_s({\bf n})$ includes only $s+\frac12$ or $s+1$ arbitrary
parameters for half integer or integer spin respectively. A
particular case of this statement has been  proven in section 2, see
equations (\ref{ps3}) and (\ref{T1}).

Thus the Hamiltonian for superintegrable model of arbitrary spin can
be represented in the following form:
\begin{gather}{\cal H}_s=\frac{p_x^2+p_y^2}{2m}+\frac1{r}{\hat\mu}_s(
{\bf n}) \label{ps211}\end{gather} where ${\hat\mu}_s( {\bf n})$ is
the matrix defined by equations (\ref{T2}) and (\ref{i}).
Hamiltonian (\ref{ps211}) is invariant w.r.t. algebra o(2,1) whose
basis elements are given by equations (\ref{J3}) and (\ref{AA}) with
${\mu}( {\bf n})\to{\hat\mu}_s( {\bf n})$.

Notice that in particular case $s=1$ Hamiltonian (\ref{ps211})
is reduced to the operator defined by equations (\ref{ps21}) and  (\ref{mus})
where $\mu=\frac12(\lambda_0-\lambda_1)$ and
$\lambda=\frac12(\lambda_0+\lambda_1)$.
For $s=3/2$ matrix (\ref{T2})
 can be transformed to the form (\ref{mu32}) with $\mu=\frac16(\lambda_{\frac32}-
\lambda_{\frac12}), \nu=\frac38\lambda_{\frac12}-\frac1{24}\lambda_{\frac32}$
by applying transformation (\ref{trans2}).

\section{Dual shape invariance and exact solutions}\label{dual}

Let us consider the eigenvalue problem (\ref{ps1}) for Hamiltonian
(\ref{ps21}), (\ref{i}). Like in section 2 we again introduce polar
coordinates (\ref{polco}) and expand solutions via eigenvectors of
operator $J_z$ (\ref{J3}) where $S_z$ is the matrix of spin $s$
defined by equation (\ref{S3}). These eigenvectors can be
represented as:
\begin{gather}\label{PSI}\psi_\kappa=\frac1{\sqrt{r}}\exp(\ri(\kappa-S_z)
\theta)\Phi_\kappa(r)\end{gather} where
\begin{gather}\label{PHI}\Phi_\kappa(r)=
\text{column}(\phi_s,\phi_{s-1},...,\phi_{-s}).\end{gather}

Substituting (\ref{ps211}), (\ref{PSI}), (\ref{PHI}) into
(\ref{ps1})  we obtain the following equation  for radial functions
$\Phi_{k}$: \beq\label{ps10}\hat{\cal
H}_\kappa\Phi_\kappa\equiv\left(-\frac{\partial^2}{\partial
r^2}+V_\kappa\right)\Phi_\kappa={\cal {\tilde E}}\Phi_\kappa\eeq
where ${\cal {\tilde E}}=\frac{\cal E}{2m}$ and
\begin{gather}
\label{HamPP1} V_\kappa=\left((k-S_z)^2-\frac14\right)
\frac{1}{r^2}+\tilde\mu_s\frac{1} {{r}},\quad \tilde\mu_s=2m\mu_s({\bf
n})|_{n_y=0}.
\end{gather}

Both the initial  equation (\ref{ps4}) and equation (\ref{ps10}) are
invariant w.r.t. the space reflection transformation (\ref{ps71})
where $\Lambda=\tilde \mu_s({\bf n})/2m\lambda$. Thus it is reasonable to
search for solutions of (\ref{ps10}) for non-negative $\kappa$ since
solutions for negative $\kappa$ can be obtained using transformation
(\ref{ps71}).

 Hamiltonian $\hat{\cal H}_\kappa$  and matrix $S_z^2$ commute between
 themselves and so they have a mutual system of eigenfunctions.
Matrix $S_z^2$ is diagonal and its eigenfunctions $\psi_\nu$
corresponding to eigenvalues $\nu^2=s^2, (s-1)^2, (s-2)^2,...$ have
two or one non-zero components for $\nu^2>0$ and $\nu=0$
respectively. In the standard representation of spin matrix with
diagonal $S_z$ and symmetric $S_x$,\footnote{See, e.g.,
\cite{Pron2}, eq. (22), or \cite{FN} eq.(4.65)} , matrix $\tilde
\mu_s$ is symmetric and antidiagonal, and has the unit non-zero
entries.

Thus the eigenvalue problem (\ref{ps10})  can be decoupled to the following
equations: \beq\label{ps100}\hat{\cal
H}_{\kappa,\nu}\psi_{\kappa,\nu}\equiv\left(-\frac{\partial^2}{\partial
r^2}+V_\kappa\right)\psi_{\kappa,\nu}={\cal {\tilde
E}}\psi_{\kappa,\nu}\eeq where
$\psi_{\kappa,\nu}=\left(\bea{c}\phi_\nu\\\phi_{-\nu}\eea\right)$
are two-component functions, and
\begin{gather}\label{Vsmall}V_{\kappa,\nu}=\frac{(\kappa-\nu \sigma_3)^2-
\frac14}{r^2}+ \frac{\tilde\lambda}{r} \sigma_1,\quad \nu\neq0.
\end{gather}

For $\nu=0$ the corresponding reduced potential is one-dimensional:
\begin{gather}\label{Vsmaller}V_{\kappa,0}=\frac{(\kappa)^2-
\frac14}{r^2}+ \frac{\tilde\lambda}{r} .
\end{gather}

 Potentials (\ref{Vsmall}) were
discussed in paper \cite{Nik1} where parameter $\nu$ was denoted as
$\mu+\frac12$. These potentials are shape invariant and can be
expressed via superpotentials as follows:
\begin{gather}
 \label{Rikikaka} V_{\kappa,\nu}=W_{\kappa,\nu}^2-
 W_{\kappa,\nu}'+c_{\kappa}\end{gather}
 where
\begin{gather}\label{SPAS1} W_{\kappa,\nu}=\frac{\nu}{r}\sigma_3-
\frac{\tilde\lambda}{2\kappa+1}\sigma_1- \frac{2\kappa+1}{2r}\end{gather}
and $c_\kappa$ is the parameter  given in equation (\ref{a+-}).

Shape invariance of $V_{\kappa,\nu}$ can be easily proven since, in
addition to the representation (\ref{Rikikaka}) the following
equation holds true:
\begin{gather}\label{SIC}V_{\kappa,\nu}^+=W_{\kappa,\nu}^2+W_{\kappa,\nu}'=
V_{k+1,\nu}+ C_\kappa\end{gather} where $C_\kappa=c_{k+1}-c_\kappa$.

One dimensional potential (\ref{Vsmaller}) is shape invariant too.
It can be represented in the standard form (\ref{Rikikaka}) with
\begin{gather}\label{SPAS3} W_{\kappa,0}=-
\frac{\tilde\lambda}{2\kappa+1}- \frac{2\kappa+1}{2r}\end{gather}

It is important to note that  potential (\ref{Vsmall}) is invariant
w.r.t. the change $\nu\to\kappa,\ \ \kappa\to\nu$, while
superpotential (\ref{SPAS1}) {\it is not invariant w.r.t. this
change}, since
\begin{gather}\label{SPAS2} W_{\nu,\kappa}=\frac{\kappa}r \sigma_3-
\frac{\tilde\lambda}{2\nu+1}\sigma_1- \frac{2\nu+1}{2r}.\end{gather}

Thus, in addition to (\ref{Rikikaka}),
 there
exist the alternative representations of potential (\ref{Vsmall})
via superpotential, i.e., equation (\ref{Rikikaka}) where
$W_{\kappa,\nu}$ is changed by superpotential (\ref{SPAS2}):
\begin{gather}\label{Rikikak} V_{\kappa,\nu}=W_{\nu,\kappa}^2-
 W_{\nu,\kappa}'+c_{\nu},\quad c_\nu=-\frac{\tilde\lambda^2}{(2\nu+1)^2}.\end{gather}
 In other words, potential (\ref{Vsmall}) appears to be shape invariant
 w.r.t. the shifts of two parameters, i.e., $\kappa$ and ${\nu}$. This is a particular case of the dual shape invariance
phenomena discovered in \cite{Nik1}.

Using representations (\ref{Rikikaka}) and (\ref{Rikikak}) we easy find
the ground state vectors of Hamiltonian (\ref{ps100}). Solving
equations (\ref{ps7}) with superpotentials (\ref{SPAS1}) and
(\ref{SPAS2}) we obtain the following components of the ground state
vectors $\psi_\nu^0=\text{column}(\phi_\nu^0,\phi_{-\nu}^0), \
\nu=s, s-1, s-2,..., \nu>0$:
\begin{gather}\phi_\nu^0=d_\nu r^{\kappa+1}
K_{\nu+\frac12}\left(\frac{\tilde\lambda r}{2\kappa+1}\right), \ \
\phi_{-\nu}^0=d_\nu(-1)^{\nu-\frac12}
r^{\kappa+1}K_{\nu-\frac12}\left(\frac{\tilde\lambda
r}{2\kappa+1}\right)\label{arb_sol},\ \ \kappa\geq\nu\end{gather}
for superpotential (\ref{SPAS1}), and
\begin{gather}\phi_\nu^0=d_\nu r^{\nu+1}K_{\kappa+\frac12}
\left(\frac{\tilde\lambda r}{2\nu+1}\right), \ \
\phi_{-\nu}^0=d_\nu(-1)^{\kappa-\frac12}
r^{\nu+1}K_{\kappa-\frac12}\left(\frac{\tilde\lambda r}{2\nu+1}\right),\ \
0\leq\kappa<\nu\label{arb_soli}\end{gather} for superpotential
(\ref{SPAS2}), were $d_\nu$ are integration constants. Solution for
$\nu=0$, i.e., the component $\phi_0^0$, is given by the following
equation:
\begin{gather}\label{gs0}\phi_0^0=du^{\kappa+\frac12}\exp\left(-\tilde\lambda r\right),\quad
\kappa=0,1,2,...
\end{gather}

Functions (\ref{arb_sol}) and (\ref{arb_soli}) are square integrable
for $\tilde\lambda>0$ and arbitrary integer or half integer $\kappa\geq 0$. The same is
true for (\ref{gs0}). However, functions (\ref{arb_sol}) for
 $\kappa\leq\nu$ and functions (\ref{arb_soli}) for
$\kappa\geq\nu$ do not vanish at $r=0$. So such values of parameters
$\nu$ and $\kappa$ should be excluded, as it is indicated in the
r.h.s. of the discussed equations.

Vectors for exited states and the corresponding energy levels again
are given by relations (\ref{ps9}), (\ref{a+-}) and
(\ref{eigenvalues}) respectively where  $W_\kappa$ should be
replaced by superpotential (\ref{SPAS1}) or (\ref{SPAS2}). In
accordance with the analysis presented in \cite{Nik1} all such
vectors are square integrable and vanish at $r=0$.

 Thus we find the eigenvectors of Hamiltonian (\ref{ps21}),
(\ref{i}) which are given by equations (\ref{PSI}), (\ref{PHI}) and
(\ref{arb_sol})--(\ref{gs0}). The corresponding solutions for
negative $\kappa$ can be easily found using transformation
(\ref{ps71}) where $\Lambda=\tilde \mu_s({\bf n})$.

 In complete
analogy with the above we can solve the eigenvalue problem for the
more general Hamiltonian (\ref{ps211}) including $[s+1]$ arbitrary
parameters $\tilde\lambda_\nu$. Actually, to this effect it is sufficient
to change $\tilde\lambda\to\tilde\lambda_\nu$ in all formulae (\ref{Vsmall}),
(\ref{SPAS1})--(\ref{arb_soli}) and (\ref{eigenvalues}).


\section{Discussion}

Shape invariant matrix potentials classified in papers \cite{Nik1} and \cite{Nik2} can appear in many realistic integrable models of quantum mechanics. In the present paper we apply the results of these papers to the interesting class of exactly solvable systems  found in \cite{Pron2}. These systems are maximally superintegrable and generalize the PS model to the case of arbitrary spin.

We find  it be interesting  to study these long-awaited generalized
models in more detail. In particular, to search for their solutions, to
examine their consistency, and  to verify weather the supersymmetry
and shape invariance of the spin $\frac12$ model are kept in  the
case of arbitrary spin. In addition, it is important to understand
the physical content of the models proposed in \cite{Pron2}. Just the
tasks enumerated above are the subjects of the present paper.

First we refine the results of paper \cite{Pron2} taking into account equivalence relations w.r.t. unitary transformations
realized by constant matrices. It is shown that up to such
equivalence  the superintegrable models for arbitrary spin $s$
include $s+1$ or $s+\frac12$ arbitrary parameters for integer or
half integer spins correspondingly, while in \cite{Pron2} these
models include $2s+1$ parameter.

The cases $s=1$ and $s=\frac32$ are considered in more detail. We
present the corresponding Hamiltonians in the form which is
convenient for physical interpretation. This interpretation is
proposed using two alternative ways. First, it is possible to treat
(\ref{ps21}) as a Hamiltonian of spin-one particle with zero charge
and zero dipole momentum, which has a non-trivial quadruple
interaction with the external field. Another possibility is to
represent the Hamiltonian  in  form (\ref{rep1}) including the
interaction term nonlinear in the external magnetic field. Analogous
alternatives exist for the models with spin $\frac32$  and with
arbitrary spin, i.e., it is possible to interpret the related
potentials as results of either multipole or non-linear interaction
of a neutral particle with an external field.

 It is shown that the considered models are shape invariant. Moreover, their effective potentials appear to be particular cases of the matrix shape invariant potentials classified in \cite{Nik1}. Using this fact it is possible to construct exact solutions of these models in a simple and straightforward way. This program has been realized in section \ref{dual} immediately for arbitrary spin.

A specific property of potentials (\ref{Vsmall}) is their dual shape
invariance, i.e., the existence of two non-equivalent corresponding
superpotentials. Exactly this property enables to find good solutions
for all combinations of eigenvalues of the spin and orbital momentum operators.

Notice that supersymmetric and superconformal aspects of  planar systems  with arbitrary spin were discussed in papers \cite{Plu1} and \cite{Plu2}. We consider another class of such systems which are chargeless, have non-trivial multipole moments, and can be integrated in closed form using their supersymmetry and shape invariance.

In the present paper the shape invariant matrix potentials
classified in \cite{Nik1} and \cite{Nik2} are used to solve
explicitly the countable set of superintegrable models. In fact the
number of exactly solvable models with matrix superpotentials some
of which have the clear physical significance is much more extended.
Among them  there are also the relativistic models discussed in
\cite{nika} and many others. We plane to present such models in the
following publications.


\begin{thebibliography}{99}
\bibitem{wint}F. Tremblay, A. V. Turbiner and P. Winternitz, J. Phys. A:
Math. Theor.  42 (2009) 242001.

\bibitem{wint2} P. Winternitz and I. Yurdusen, J. Math. Phys. 47(2006) 103509; \\
P. Winternitz and I. Yurdusen, J. Phys. A: Math. Theor. 42 (2009) 385203.

\bibitem{Gen} L. Gendenshtein, JETP Lett. 38 (1983) 356.


\bibitem{Pron}G. P. Pron'ko, Y. G. Stroganov, Sov. Phys. JETP 45 (1977)
1075.

\bibitem {Vor}A. I. Voronin, Phys. Rev. A 43 (1991) 29.

\bibitem {Gol} L. V. Hau, G. A. Golovchenko, and M. M. Burns, Phys.
Rev. Lett. 74 (1995) 3138.

\bibitem{Blum1}R. Bl\"umel and K. Dietrich, Phys. Lett. A 139 (1989) 236.

\bibitem{Blum2}R. Bl\"umel and K. Dietrich, Phys. Rev. A 43 (1991) 22.

\bibitem{mota}D. Martinez, V.D. Granados and R.D. Mota, Phys. Lett. A
350 (2006) 31.

\bibitem{rodric}R. de Lima Rodrigues,  V. B. Bezerra and A. N. Vaidyac, Phys. Lett. A
287 (2001) 45.

\bibitem{Ioffe} M. V. Ioffe, S. Kuru, J. Negro and L. M. Nieto,
J. Phys. A: Math. Theor. 39 (2006) 6987.

\bibitem{nika} E. Ferraro, N. Messina and A. G. Nikitin,
          Phys. Rev. A 81 (2010) 042108.

\bibitem{Gol2} L.Vestergaard Hau, J.A. Golovchenko and M.M. Burns,
Phys. Rev. Lett. 74 (1995) 3138; \\ K. Berg-Sorensen, M. M. Burns,
J. A. Golovchenko and L. Vestergaard Hau, Phys. Rev. A 53 (1996)
1653.

\bibitem{Pron2} G. P. Pronko,  J. Phys. A: Math. Theor. 40 (2007) 13331.

\bibitem{Beckers} J. Beckers, N. Debergh, and A. G. Nikitin,
 Fortsch. der Phys. 43 (1995) 81.

 \bibitem{Nik1} A. G. Nikitin and Y. Karadzhov, J. Phys. A:
44 (2011) 305204.

 \bibitem{Nik2}A. G. Nikitin and Y. Karadzhov,  J. Phys. A: 44 (2011)
 445202.

 \bibitem{NKU}A. G. Nikitin and O. Kuriksha, Commun Nonlinear
Sci Numer Simulat (2012),
http://dx.doi.org/10.1016/j.cnsns.2012.04.009

  \bibitem{FGN} W. I. Fushchich, A. L. Grishchenko and A. G. Nikitin, Theor. Math. Phys. 8 (1971) 766.

  \bibitem{FN}  W.I. Fushchich and A. G. Nikitin, Symmetries of Equations of
     Quantum Mechanics. N.Y., Allerton Press Inc., 1994.

  \bibitem{Plu1} S. M. Klishevich and  M.S. Plyushchay,
 Nucl. Phys. B 616 (2001) 403.

 \bibitem{Plu2} A. Anabalon and M. S. Plyushchay,
 Phys. Lett. B 572 (2003) 202.

\end{thebibliography}
\end{document}